\documentclass[twocolumn,preprintnumbers,amsmath,amssymb,superscriptaddress]{revtex4}%
\usepackage{graphicx}
\usepackage{dcolumn}
\usepackage{bm}
\usepackage{amsmath}
\usepackage{amsfonts}
\usepackage{amssymb}
\usepackage{color}%
\begin{document}
\title{Detection of small single-cycle signals by stochastic resonance using a bistable
superconducting quantum interference device}
\author{Guozhu Sun}
\email{gzsun@nju.edu.cn}
\affiliation{National Laboratory of Solid State Microstructures and Research Institute of
Superconductor Electronics,School of Electronic Science and Engineering,
Nanjing University, Nanjing 210093, China}
\affiliation{Synergetic Innovation Center of Quantum Information and Quantum Physics,
University of Science and Technology of China, Hefei, Anhui 230026, China}
\author{Jiquan Zhai}
\affiliation{National Laboratory of Solid State Microstructures and Research Institute of
Superconductor Electronics,School of Electronic Science and Engineering,
Nanjing University, Nanjing 210093, China}
\affiliation{Synergetic Innovation Center of Quantum Information and Quantum Physics,
University of Science and Technology of China, Hefei, Anhui 230026, China}
\author{Xueda Wen}
\affiliation{Department of Physics, University of Illinois at Urbana-Champaign, Urbana, IL
61801, USA}
\author{Yang Yu}
\affiliation{School of Physics,
Nanjing University, Nanjing 210093, China}
\affiliation{Synergetic Innovation Center of Quantum Information and Quantum Physics,
University of Science and Technology of China, Hefei, Anhui 230026, China}
\author{Lin Kang}
\affiliation{National Laboratory of Solid State Microstructures and Research Institute of
Superconductor Electronics,School of Electronic Science and Engineering,
Nanjing University, Nanjing 210093, China}
\affiliation{Synergetic Innovation Center of Quantum Information and Quantum Physics,
University of Science and Technology of China, Hefei, Anhui 230026, China}
\author{Weiwei Xu}
\affiliation{National Laboratory of Solid State Microstructures and Research Institute of
Superconductor Electronics,School of Electronic Science and Engineering,
Nanjing University, Nanjing 210093, China}
\affiliation{Synergetic Innovation Center of Quantum Information and Quantum Physics,
University of Science and Technology of China, Hefei, Anhui 230026, China}
\author{Jian Chen}
\affiliation{National Laboratory of Solid State Microstructures and Research Institute of
Superconductor Electronics,School of Electronic Science and Engineering,
Nanjing University, Nanjing 210093, China}
\author{Peiheng Wu}
\affiliation{National Laboratory of Solid State Microstructures and Research Institute of
Superconductor Electronics,School of Electronic Science and Engineering,
Nanjing University, Nanjing 210093, China}
\affiliation{Synergetic Innovation Center of Quantum Information and Quantum Physics,
University of Science and Technology of China, Hefei, Anhui 230026, China}
\author{Siyuan Han}
\affiliation{Department of Physics and Astronomy, University of Kansas, Lawrence, KS 66045, USA}

\begin{abstract}
We propose and experimentally demonstrate detecting small single-cycle and
few-cycle signals by using the symmetric double-well potential of a radio
frequency superconducting quantum interference device (rf-SQUID). We show that
the response of this bistable system to single- and few-cycle signals has a non-monotonic dependence
on the noise strength. The response, measured by the probability of transition from initial potential well to the opposite one, becomes maximum when the noise-induced
transition rate between the two stable states of the rf-SQUID is comparable to
the signal frequency. Comparison to numerical simulations shows that the
phenomenon is a manifestation of stochastic resonance.

\end{abstract}
\maketitle

It is a long-held belief that noise is detrimental or even destructive to detecting signals which often appear as weak periodic modulations. However, during the last 35 years theoretical and experimental investigations have shown that in
nonlinear systems a proper amount of noise can actually increase the signal-to-noise ratio (SNR) and thus become beneficial for signal detection. This interesting phenomenon is named as stochastic resonance (SR) \cite{Nature.373.33,RevModPhys.70.223,IEEE997828,RPP}.
For example, suppose that a particle is moving in a periodically
perturbed symmetric double-well potential under the influence of a
Gaussian-white noise such as thermal fluctuation. Then SNR of the power spectral density of the particle's trajectory displays a broad maximum when the rate of inter-well transitions, which depends on noise strength exponentially, is comparable to the frequency of periodic signal. This is the essence of SR.

Due to its simplicity and ubiquity of the underlying mechanism, SR has attracted much
interest from physicists, chemists, biologists, and electronic engineers \cite{Nature.373.33,RevModPhys.70.223,IEEE997828,RPP,PhysRevLett.72.1947,PhysRevLett.76.1611,PhysRevLett.98.170601,APL1.2430689,APL1.3556314,APL1.3600329,APL1.4766946}. It has also been observed in Josephson junction-based systems \cite{APL1995,JAP,LTP,PhysRevE.75.021107,PhysRevE.79.030104}, which have recently attracted much interest and been applied in many fields such as quantum information \cite{PhysicsToday.58.11,Nature.474.589,RevModPhys.85.623}. However SR has been only investigated for periodic signals that last many cycles. Namely, only the steady-state
properties of the noisy periodically driven systems have been studied. On the
other hand, in a variety of science and engineering disciplines, it is a
significant challenge to detect small signals which not only last a few cycles
but also are buried in noise. Up to this point, whether SR can also enhance single-cycle signal detection remains an open question.

In this Letter, we report on the observation of SR in a radio frequency superconducting quantum
interference device's (rf-SQUID's) \cite{PhysRevLett.75.1614,apl94} response
to weak single-cycle and few-cycle signals by measuring the inter-well transition probability as a
function of the noise strength $D$ and the signal frequency $f_{s}$ systematically. Our
experimental and numerical results show that one can distill small
single-cycle and few-cycle signals from noisy environment by using bistable systems
configured as binary threshold detectors. The maximum sensitivity is
achieved at the value of $D$ that matches well with the position of SR. We
also show that the sensitivity of detecting single-cycle signals is similar to
that of detecting many-cycle signals.

In our experiment we use an rf-SQUID, which is a superconducting loop of inductance
$L$ interrupted by a Josephson junction of critical current $I_{c},$ as our
bistable detector. An optical micrograph of the sample is shown in the inset
of Fig. 1. The Josephson junction is made of Nb/AlOx/Nb on a silicon
substrate. The critical current $I_{c}$ and the capacitance $C$ of the
junction are approximately $0.80$ $\mu$A and $90$ fF, respectively. The
inductance $L$ of the Nb superconducting loop is approximately $1053$ pH. The
potential energy of the rf-SQUID is given by
\begin{equation}
U(\Phi)=\frac{(\Phi-\Phi_{e})}{2L}^{2}-E_{J}\cos\left(  \frac{2\pi\Phi}%
{\Phi_{0}}\right),
\end{equation}
where $\Phi_{0}$ is the flux quantum and $E_{J}=I_{c}\Phi_{0}/2\pi$ is the
Josephson coupling energy of the junction. The shape of the double well
potential can be controlled \emph{in situ} by a flux bias $\Phi_{e}$ applied
via a flux bias line coupled inductively to the rf-SQUID. In particular, at
$\Phi_{e}=\Phi_{0}/2$ the SQUID has a symmetric double-well potential
separated by a barrier $\Delta U_{0}$ as shown in Fig. 1. For the SQUID
studied here, $\Delta U_{0}/k_{B}\simeq12.3$ K, where $k_{B}$ is the Boltzmann's constant. The dynamics of the rf-SQUID,
identical to that of a fictitious flux particle of mass $C$ moving in the
potential $U(\Phi)$ with a friction coefficient $R^{-1},$ is governed by the
corresponding Langevin equation:
\begin{equation}
C\frac{d^{2}\Phi}{dt^{2}}+\frac{1}{R}\frac{d\Phi}{dt}=-\frac{dU}{d\Phi}%
+I_{n}(t). \label{EOM}%
\end{equation}
Here, $I_{n}$ is the noise current and $R$ is the damping resistance of the
Josephson junction. Without externally injected noise, $I_{n}$ and $R$ are
related by the fluctuation-dissipation theorem $\langle I_{n}(t)I_{n}%
(t^{^{\prime}})\rangle=\frac{2k_{B}T}{R}\delta(t-t^{^{\prime}})$ in thermal
equilibrium, where $T$ is temperature. The small oscillation frequency of the system around the bottom
of the potential wells is denoted as $\omega_{0}.$ At $T\gg\hbar\omega
_{0}/k_{B},$ where $\hbar$ is the Planck constant, thermal activation causes
inter-well hopping with the characteristic transition rate given by \cite{RevModPhys.62.251}%
\begin{equation}
\Gamma_{0}=\frac{\omega_{0}}{2\pi}a_{t}\exp\left(  -\frac{\Delta U_{0}}%
{k_{B}T}\right)  , \label{rate0}%
\end{equation}
where $a_{t}$ is a damping dependent constant of order of unity. When transitions are dominated by an external noise source of strength $D\gg
k_{B}T$, the denominator in the exponent of Eq. (\ref{rate0}) is replaced by
$D$ which is proportional to the square of the rms noise current,
$I_{n,\text{rms}}^{2},$ applied to the system. \ For the sake of simplicity,
hereafter we set $k_{B}=1$ so that $D$ is measured in units of kelvin. \ Note
that because the potential is symmetric, $\Gamma_{0}$ is identical for
left-to-right and right-to-left transitions.

Because all key parameters of the rf-SQUID potential and its control circuit
can be accurately determined, the double-well potential of the rf-SQUID is an
ideal system for investigating SR
\cite{APL1995,JAP,LTP} and
noise-enhanced detection of single-cycle and few-cycle signals. In our experiment,
the Gaussian-white noise has a bandwidth of about $9$ MHz, which is generated
by an arbitrary waveform generator. The signal and noise are applied to the
rf-SQUID through a second flux bias line with higher bandwidth (up to $18$
GHz). The relationship between $D$ and $I_{n,\text{rms}}^{2}$ is calibrated by
measuring $\Gamma_{0}$ versus $I_{n,\text{rms}}^{2}$ and comparing the result to
Eq. (\ref{rate0}).

As shown schematically in Fig. 1, each measurement cycle begins by ramping up
the quasi-static flux bias from $0$ to $\Phi_{0}/2$ to prepare the flux
particle in the left side of the symmetric double-well potential. This is
followed by applying a single-cycle modulation of flux bias that causes the
potential barrier to oscillate as $\Delta U_{0}+\varepsilon_{0}\sin(2\pi
f_{s}t)$, where $\varepsilon_{0}$ is proportional to the amplitude of the flux
modulation which is kept at $\varepsilon_{0}=0.07\Delta U_{0}\simeq0.86$ K in
the experiment. The position of the flux particle is measured by using a
dc-SQUID switching magnetometer inductively coupled to the rf-SQUID, either
after a single signal cycle or a fixed duration of signal time as discussed later, as a function
of $0.5$ K$\leq D\leq2.0$ K and $10$ kHz $\leq f_{s}\leq200$ kHz. The
quasi-static flux bias is then ramped down to zero to complete the measurement
cycle. To obtain the fractional population in the right well $\rho_{R}$, the
procedure is repeated $2000$ times at each value of $D$ and $f_{s}$. All data
are measured at $T\approx20$ mK $\ll D$ in a cryogen-free dilution fridge
carefully shielded from the environmental electromagnetic interference so that the
effects of thermal fluctuation and extra noise on the experiment are negligible.

We first measure $\rho_{R}$ as a function of the noise strength $D$ by using
single-cycle signals $\varepsilon_{\pm}(t)=\pm\varepsilon_{0}\sin(2\pi
f_{s}t)$ as depicted in Fig. 1, where the signal frequency $f_{s}=10$ kHz.
The result is shown in Fig. 2(a). The noise strength $D$ is varied
between $0.5$ K and $2.0$ K. Therefore, transitions between the two potential
wells at $\varepsilon_{0}=0$ (no signal) are noise activated. Note that
without the noise, $\varepsilon(t)$ alone would be too small to cause
transitions between the potential wells because $(\Delta U_{0}-\varepsilon
_{0})/T>500$. Hence, $D>0.5$ K is required for the flux particle to hop from
one well to the other within the duration of each signal cycle. On the other
hand, at $\varepsilon_{0}=0$ the transition rate $\Gamma_{0}$ grows
exponentially from approximately $1/$s at $D=0.5$ K to greater than $10^{7}/$s
at $D=2.0$ K, as shown in the inset of Fig. 2(a). It can be seen that for $D\lesssim0.7$ K the population of the
right well $\rho_{R}$ is negligible at $t=\tau=1/f_{s}$. The data indicated by the blue squares in Fig. 2(a) are taken with $\varepsilon_{+}(t)$ which show that as $D$ is increased
from $0.7$ K, $\rho_{R}$ rises rapidly to reach maximum when $\Gamma
_{0}(D_{\text{m}})\sim 2f_{s}$, where $D_{\text{m}}$ denotes the noise
strength corresponding to the maximum $\rho_{R}$. When $D>$ $D_{\text{m}},$
the probability of hopping back from the right well to the left well increases rapidly, causing $\rho_{R}(D)$ to decrease. Finally, when $D\gg$ $\varepsilon
_{0}$ the population of each potential well is equalized to $50\%$.

SR has two most prominent signatures: One is the peak in the system's response
versus noise strength $D$. The other is the position of the peak and the
signal frequency satisfying $\Gamma_{0}(D_{\text{m}})\sim 2f_{s},$ or
equivalently, $1/D_{\text{m}}\sim-\ln(f_{s})$ according to Eq. (\ref{rate0}).
In Fig. 2(b), where $\varepsilon_{+}(t)$ is applied, we plot $\rho_{R}$ versus $1/D$ and $f_{s}$ which shows clearly both signatures of SR. In particular, the nearly linear relationship between $\Gamma_{0}(D_{\text{m}})$ and $f_{s}$ is demonstrated as shown in the inset of Fig. 2(b). The slope of $\Gamma_{0}(D_{\text{m}})$ versus $f_{s}$ obtained from the best-fit to a line is 2.7, which is consistent with the numerical result previously obtained for SR under continuous modulation \cite{PhysRevE.48.3390}. In addition, we numerically calculate the power
spectral density $S(f_{s})$ of the flux particle's trajectories $\Phi(t)$
generated by Monte Carlo simulation of Eq. (\ref{EOM}). It is found that
$S(f_{s})$ reaches its maximum at the same value of $D$ as $\rho_{R}$ does. We
thus conclude that SR plays a central role in the bistable system's response
to single-cycle signals.

In order to compare the result of our measurements with that of numerical
study over the entire parameter space covered by the experiment, we adopt the
two-state model \cite{RPP,PhysRevA.39.4854} and introduce the rate equation:%

\begin{equation}
\frac{d\rho_{R}(t)}{dt}=-\Gamma_{-}\rho_{R}(t)+\Gamma_{+}[1-\rho_{R}(t)] \label{population}
\end{equation}
with the initial condition $\rho_{L}(0)=1,$ $\rho_{R}(0)=0.$ Here, $\rho_{R}$
($\rho_{L}=1-\rho_{R}$) is the fractional population of the right (left)
potential well. When $\varepsilon_{0}\neq0$, the barrier height is oscillating
between $\Delta U_{0}\pm\varepsilon_{0}$ and the transition rates are time-dependent%
\begin{equation}
\Gamma_{\pm}(t)=\Gamma_{0}\exp\left[  -\frac{\pm\varepsilon_{0}\sin(2\pi
ft)}{D}\right]  , \label{rates}%
\end{equation}
where $\Gamma_{+}$ and $\Gamma_{-}$ denote the rates of left-to-right and
right-to-left transitions, respectively. $\Gamma_{0}$ is given by Eq.
(\ref{rate0}). Using the system parameters given above, we numerically
integrate Eq.(\ref{population}) to obtain $\rho_{R}(t)$ as a function of $f_{s}$ and
$D.$ The result is shown in Fig. 2(c). \ It can be seen that the
key features of the experimental data are well reproduced.

Next, we show that the sensitivity of detecting single-cycle signals is
comparable to that of detecting many-cycle signals and that one can predict the
population distribution of the bistable systems at the end of $N$-cycle
modulations $\rho_{R,N}\equiv\rho_{R}(N\tau)=1-\rho_{L,N}$ from that of
single-cycle modulation $\rho_{R,1}$. It is
straightforward to obtain the recursion relation%
\begin{align}
\rho_{R,n+1}  &  =\rho_{L,n}P_{+}+\rho_{R,n}(1-P_{-})\label{recur}\\
&  =(1-\rho_{R,n})P_{+}+\rho_{R,n}(1-P_{-}).\nonumber
\end{align}
The first (second) term of the r.h.s. of Eq. (\ref{recur}) is the
fractional population of the left (right) well at $t=n\tau$ that ends
(remains) in the right well at $t=(n+1)\tau$. $P_{+(-)}$ is the
probability of switching from the left (right) to the right (left) well during
the time interval $n\tau\leq t\leq(n+1)\tau$. Notice that with the
single-cycle perturbation $\varepsilon_{\pm}(0\leq t\leq\tau)$ and the initial
condition $\rho_{R,0}=0$, one has$\ P_{\pm}=\rho_{R,1}$ by taking into
consideration the spatial and temporal symmetry properties of the rf-SQUID
potential and $\varepsilon_{\pm}$. Thus, we can obtain $P_{\pm}$ directly
from the data presented in Fig. 2(a). Because Eq.
(\ref{recur}) is valid for arbitrary noise strength $D$ and signal frequency
$f_{s}$, we can compute $\rho_{R,N}$ from $P_{\pm}$ for any integer $N>1$. We
find that as $N$ increases $\rho_{R,N}$ converges rapidly. In order to
investigate the dependence of SR on the number of signal cycles $N$, we modify
the experimental procedure by changing the duration of the applied signal and
noise from $\tau$ to $0.3$ ms. Thus, we have $N=3$ for $f_{s}=10$ kHz, which
increases ultimately to $N=60$ for $f_{s}=200$ kHz. In Fig. 3(a), the measured
$\rho_{R,N}$ is plotted against $1/D$ and $f_{s}$, which compares
well with $\rho_{R,N}$ computed from Eq. (\ref{recur}) by using the measured
$P_{\pm}$ as inputs [see Fig. 3(b)] and that obtained by solving the
corresponding rate equation (\ref{population}) [see Fig. 3(c)]. The
results presented in Fig. 3 all have two distinctive features: (i) The
threshold noise strength $D_{0}$ which demarcates the blue region
($\rho_{R,N}\approx0)$ and the yellow region depends weakly on the number of
signal cycles $N,$ and (ii) $1/D_{\text{m}}\varpropto-\ln(f_{s})$ remains
valid for the entire range of $3\leq N\leq60$. As shown in the inset of Fig. 3(a), the dependance of $\Gamma_{0}(D_{\text{m}})$ on $f_{s}$ is approximately linear with a slope of about 2.6. These two features strongly indicate that the sensitivity of detecting single-cycle signals is similar to that of many-cycle and continuous wave signals and that SR does exist in the systems driven by small single-cycle signals.

In summary, using an rf-SQUID as a prototypical bistable system, we have
demonstrated the existence of SR with single-cycle perturbation to the
symmetric double-well potential of the system. Furthermore, we have
investigated the possibility of exploiting SR for detecting
small single-cycle and few-cycle signals in noisy environment. We have found
that a proper amount of noise can lead to SR which enhances the sensitivity of
detection. Our work provides insights into the behavior of bistable systems
under the combined influence of weak single-cycle (or few-cycle) periodic modulation
and noise. Because conventional techniques, such as phase sensitive
lock-in and heterodyne detection schemes, are not applicable to detecting
single-cycle and few-cycle signals buried in noise, the method demonstrated
here is promising for applications where signals are unavoidably mixed up with
noise and only last a very small number of cycles.

We thank Dan-Wei Zhang and Shi-Liang Zhu for the valuable discussions. This
work was partially supported by MOST (Grant Nos. 2011CB922104 and 2011CBA00200), NSFC (11474154, BK2012013), PAPD, a doctoral program
(20120091110030) and Dengfeng Project B of Nanjing University. S.H. was
supported in part by NSF (PHY-1314861).


\newpage

\begin{figure}[ptb]
\includegraphics[width=3.5in]{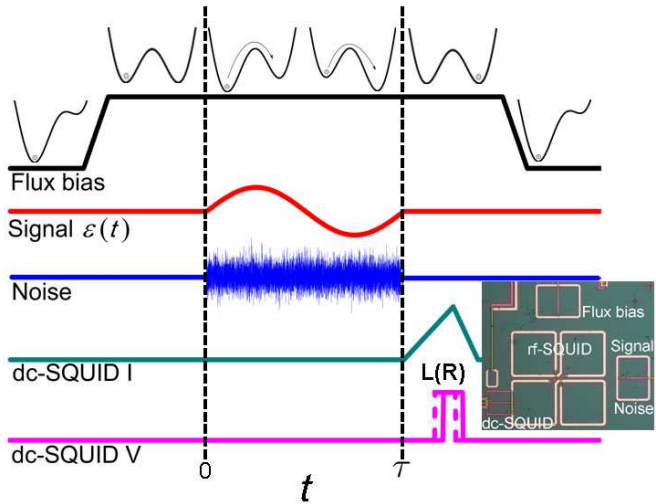}
\caption{A time profile of manipulation and measurement. Potential wells
at several key moments are also plotted. The inset shows an optical
micrograph of a Nb/AlOx/Nb rf-SQUID with an inductively coupled dc-SQUID and
flux bias lines.}%
\label{fig:epsart}%
\end{figure}

\begin{figure}[ptb]
\includegraphics[width=3.5in]{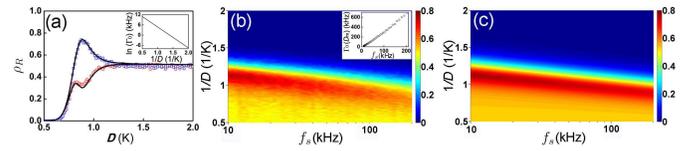}
\caption{\textbf{(a)} Measured $\rho_{R}(D)$ with a single-cycle sinusoidal
signal. $f_{s}$ = 10 kHz. Data indicated by the blue squares and red circles
correspond to $\varepsilon_{+}(t)$ and $\varepsilon_{-}(t)$, respectively,
which agree with the numerical calculation (black lines). The inset shows the linear dependance of ln$(\Gamma_{0})$ on $1/D$. \textbf{(b)}
$\rho_{R}$ as a function of $1/D$ and $f_{s}$ with single-cycle sinusoidal
signals $\varepsilon_{+}(t)=\varepsilon_{0}\sin(2\pi f_{s}t)$. As shown in the inset, when $\rho_{R}$ reaches a maximum,
$\Gamma_{0}(D_{\text{m}})\sim 2f_{s}$, where $D_{\text{m}}$
denotes the noise strength corresponding to maximum $\rho_{R}$. These
results are consistent with the theoretical hypothesis of SR. \textbf{(c)} Numerical calculation of $\rho_{R}$ as a function of $1/D$ and $f_{s}$ with single-cycle sinusoidal
signals $\varepsilon_{+}(t)=\varepsilon_{0}\sin(2\pi f_{s}t)$, which agree well with the experimental data.}%
\label{fig:epsart}%
\end{figure}

\begin{figure}[ptb]
\includegraphics[width=3.5in]{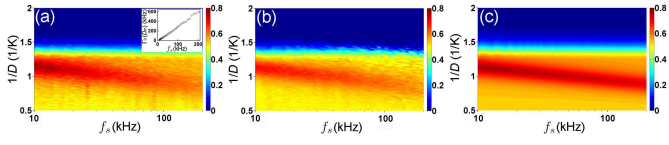}
\caption{\textbf{(a)} Measured $\rho_{R}$ as a function of $1/D$ and $f_{s}$
with a constant duration ($0.3$ ms). SR remains as shown in the inset.
\textbf{(b)} $\rho_{R}$ as a function of $1/D$ and $f_{s}$ with a constant duration ($0.3$ ms), derived from the recursion relation Eq. (\ref{recur}) and
$P_{\pm}$ obtained from the experimental results when single-cycle signals are
used. \textbf{(c)} Numerical calculation of $\rho_{R}$ as a function of $1/D$ and $f_{s}$
with a constant duration ($0.3$ ms).}%
\label{fig:epsart}%
\end{figure}

\end{document}